\documentclass[12pt]{article}
\usepackage{amssymb}

\newcommand{\cA}{{\cal A}}

\newcommand{\cK}{{\cal K}}

\newcommand{\cM}{{\cal M}}
\newcommand{\cN}{{\cal N}}

\newcommand{\cT}{{\cal T}}

\newcommand{\cZ}{{\cal Z}}
\newcommand{\bo}{{\bar o}}

\def\a{\alpha}

\def\i{{\rm i}}

\def\s{\sigma}

\def\G{\Gamma}

\def\O{\Omega}

\def\hlf{{1\over 2}}
\def\ZZ{\mZ}
\def\bo{{\raise.15ex\hbox{\large$\Box$}}}               

\def\face{{\raise.2ex\hbox{$\displaystyle \bigodot$}\mskip-2.2mu \llap {$\ddot
        \smile$}}}                                      

\def\leftrightarrowfill{$\mathsurround=0pt \mathord\leftarrow \mkern-6mu
        \cleaders\hbox{$\mkern-2mu \mathord- \mkern-2mu$}\hfill
        \mkern-6mu \mathord\rightarrow$}       
\def\dvec#1{\vbox{\ialign{##\crcr
        \leftrightarrowfill\crcr\noalign{\kern-1pt\nointerlineskip}
        $\hfil\displaystyle{#1}\hfil$\crcr}}}           

\def\beq{\begin{equation}}
\def\eeq{\end{equation}}

\def\beqx{\begin{displaymath}}
\def\eeqx{\end{displaymath}}

\def\beql{\begin{eqnarray}}
\def\eeql{\end{eqnarray}}

\newcommand{\bea}{\begin{eqnarray}}
\newcommand{\eea}{\end{eqnarray}}

\newcommand{\R}[1]{(\ref{eq:#1})}

\def\[{\left [}
\def\]{\right ]}
\def\({\left (}
\def\){\right )}

\def\ZZ{\mathbb{Z}}

\def\cA{{\cal A}}

 \def\cK{{\cal K}} 
\def\cM{{\cal M}} \def\cN{{\cal N}} 
  
 \def\cT{{\cal T}} 
  
 \def\cZ{{\cal Z}}

\def\hlf{\frac{1}{2}}

\def\+{\oplus}

\begin{document}

\vspace*{0.15in}

\begin{center}
{\Large \bf Comments on the classification of orientifolds\footnote{Based on a
seminar given at the RTN-workshop in Leuven, Belgium, September 2002.}}

\vspace*{0.5in}
{L. R. Huiszoon}\footnote{email:lennaert@itf.fys.leuven.ac.be}\\[.3in]
{\em  Instituut voor Theoretische Fysica\\
Katholieke Universiteit Leuven\\
 Celestijnenlaan 200D\\
B-3001 Leuven}\\[0.2in]

\end{center}

\begin{center}
{\bf
Abstract}

The simple current construction
 of orientifolds
based on rational conformal field theories is reviewed.
When applied to $SO(16)$ level $1$,
 one can describe all ten-dimensional
orientifolds in a unified framework.

\end{center}

\section{Introduction}

It is well known that M-theory has an infinite number of vacua.
 As long as we do not know the
vacuum selection principle of M-theory, it might be a good strategy
 to study generic properties of these vacua. A large class of vacua are string
 compactifications with D-branes and O-planes. 
Such {\em orientifolds} have been extensively studied in flat space and recently
in curved spaces~\cite{BlBKL}.
 In this talk I will present a general
prescription~\cite{FOE} to construct orientifolds that are
 based on rational conformal field theories (RCFT). The common property of
  these theories is that the RCFT has simple currents~\cite{simple} which enable us to
  find universal formulas for D-branes and O-planes.   

The outline of this talk is as follows. I will explain what simple currents are 
and how they can be used to construct string theories. 
From time to time I will clarify this construction with an
 example, the bosonic string compactified on the $E_8\times SO(16)$ root lattice\footnote{
 The `dummy' $E_8$ will de dropped in the remainder.}.
This example is interesting in its own right, since suitable extensions and truncations
of this theory are related to the $d=10$ fermionic strings~\cite{CaENT,BSM,ChEHT}. We will see that
our prescription describes all known $d=10$ orientifolds~\cite{prop} in one unified framework.

\section{Closed strings and simple currents}

The worldsheet of a closed oriented string must be a conformal field theory and splits in
a left- and rightmoving sector.
 The chiral half of
 these theories is specified by a (chiral) algebra $\cA$, that we assume to be rational.
  Rationality simply means that the number of
 (chiral) primaries of $\cA$, labelled by $m$, is finite.
   Let $Z_{mn}$ denote how many times the left-moving  primary $m$ is
  combined with the right-moving primary $n$. Then we must have
  \beq
  [Z,S]=[Z,T] = 0
  \eeq
  in order for this theory to be finite. Here $S$ and $T$ are the
  modular matrices of $\cA$. The matrix $Z$ is refered to as the {\em
  modular invariant}. The first step will be to classify
  these. This can (to a certain extent) be done by simple currents~\cite{simple,KrS}.

  {\em Simple currents} are primary fields $J$ whose fusion product with any
  primary $m$ yields just one primary $J \times m \equiv Jm$.
  An illustrative example is the chiral algebra of
  free bosons. For this theory, the
  primaries are labeled by momenta $p$ and the fusion product expresses momentum
  conservation $p \times q
  = p+q$. So every primary is a simple current. This example
  falls strictly speaking not in the class we are considering since the chiral
  algebra is not rational. However, when
   compactified on a circle of rational radius, the
  chiral algebra can be made rational. We can for instance compactify
  $r$ bosons on the coroot-lattice of a Lie group $G$ of rank $r$.
  The momenta are $G$ weights,
  but weights that differ by a root can be grouped into a single primary of an
  `extended' algebra called $G$ level $1$.
   The number of primaries of
  this extended algebra therefore equals the number
  of conjugacy classes of $G$. Since this theory still consists of free bosons,
    all primaries are
   simple currents. The fusion rules reflect momentum conservation modulo roots
  which yields a group structure that is isomorphic to the centre of $G$.

   Throughout this
 talk we will simplify the discussion in two ways.
  First, the simple current group
 that generates the modular invariant is $\ZZ_N$. Second, we assume
 that these currents do not have fixed
 points\footnote{For the general case including fixed points, see~\cite{FOE}.}.
 Such a
 cyclic group defines a modular invariant whose only nonzero entries are
 $Z_{m,J^\a m}=1$ and this happens only when
 $$h_m+(1-\a)h_J-h_{Jm} \in \ZZ \;\; .$$
  Here $J$ is the generator of
 $\ZZ_N$. We have $J^N=0$, the vacuum. The index $\a$ runs from $1$ to $N$ and $h_m$
 is the conformal weight of $m$.

 Consider $SO(16)$ level $1$. The primaries
  $(O_{16},S_{16},V_{16},C_{16})$  have conformal weights $(0,1,1/2,1)$
  and generate a $\ZZ_2\times \ZZ_2$ simple current group.
 The $\ZZ_2$ simple current invariants corresponding to $O_{16},V_{16},S_{16}$ are\footnote{The invariant 
 obtained from $C_{16}$ gives after the truncation~\R{PSSR} a fermionic 
 spectrum that is equivalent to that of $\cZ_{IIB}$.}
 :
 \bea
 \cZ_{0B}  & = & O_{16}\bar{O}_{16} + V_{16}\bar{V}_{16}+ C_{16}\bar{C}_{16} + S_{16}\bar{S}_{16} \;\;\;,\\
 \cZ_{0A}  & = & O_{16}\bar{O}_{16} + V_{16}\bar{V}_{16} + C_{16}\bar{S}_{16} + S_{16}\bar{C}_{16}
 \;\;\; ,\\
 \cZ_{IIB} & = & (O_{16}+S_{16})(\bar{O}_{16} + \bar{S}_{16}) \;\;\;.
\eea
Note that we have written $\cZ= \sum_{mn}\chi_m Z_{mn} \bar{\chi}_n$ where
$\chi_m$ is the Virasoro specialized character of primary $m$. We
denote the characters of $SO(16)$ by $O_{16} \equiv \chi_{O_{16}}$ etcetera. Right-moving
characters are barred.
The spectra of these bosonic theories
can be related to the fermionic string theories as indicated by the subscript.
In order to read off the fermionic spectrum, one has to replace the $SO(16)$ characters by $SO(8)$
characters as follows~\cite{BSM}:
\beq \label{eq:PSSR}
O_{16} \rightarrow V_8 \;\;\; , \;\;\; V_{16} \rightarrow O_8 \;\;\; , \;\;\;
S_{16} \rightarrow -S_8 \;\;\; , \;\;\; C_{16} \rightarrow -C_8 \;\;\; .
\eeq
The $O_{16}, V_{16}$ are in the NS sector and $S_{16}, C_{16}$ are in the R sector. Due to the minus
signs in~\R{PSSR}, states in the $RNS$ and $NSR$ sectors contribute with a minus sign to the partition
function and thus describe fermions. Type OB and OA do not contain fermions, are not space-time
supersymmetric and are tachyonic due to the groundstate of $V_{16}\bar{V}_{16}$. Type IIB
is supersymmetric since it contains gravitinos from $S_{16}\bar{O}_{16}$ and $\bar{S}_{16}O_{16}$
 and is tachyon free.

When we use the full $\ZZ_2\times \ZZ_2$ centre, the resulting invariant describes the type IIA
string:
\beq
 \cZ_{IIA} = (O_{16}+S_{16})(\bar{O}_{16} + \bar{C}_{16}) \;\;\; .
\eeq
Since this theory is not invariant under $\O$, we cannot perform the standard orientifolding. However,
this theory is invariant under $\cT^{-1} \O \cT$ where $\cT$ interchanges
$\bar{S}_{16}$ and $\bar{C}_{16}$.
 Orientifolds of
type IIA are beyond the scope of this talk.

From this example we see
that all uncompactified string theories are simple current invariants. For
a generic
compactified string theory this is no longer true, since the `internal' CFT might be
interacting (i.e., is not described by free bosons). For interacting
CFT's, not all primaries are simple currents, but many invariants are
simple current invariants. Note from our example
that simple currents nicely
implement the ten dimensional GSO projection;
 this is true for any dimension~(see for instance~\cite{FSW}.

\section{Orientifolds and simple currents}

Next we consider two-dimensional CFT's on worldsheets with 
boundaries and crosscaps~\cite{cardy, Sa, BiSa, SaSt, AnSa}.
 A necessary
ingredient for calculating the
correlation functions on such surfaces are the one-point functions of closed strings on the disc and
$RP^2$, i.e., the {\em tadpoles}. Let $|\phi_{ij}\rangle$ denote a closed oriented string state whose
left/right-mover is (a descendant of) the primary $i/j$. Then the disc and crosscap tadpoles of this field
are given by
\beq
\langle B |  \phi_{ij} \rangle \;\;\; , \;\;\; \langle C |  \phi_{ij} \rangle  \;\;\;.
\eeq
Here $|B\rangle$ and $|C\rangle$ are {\em boundary} and {\em crosscap states} whose presice definition is not
important for the rest of this talk. These tadpoles are constrained as follows. First note that on a surface with
a boundary, the left- and rightmovers on the string are no longer independent. In
particular, the left- and rightmoving worldsheet currents are related by a boundary condition. We can for
instance choose a diagonal `gluing condition' in which the left- and rightmoving currents are equal. Then, the
full $\cA \times \bar{\cA}$ algebra is broken to a diagonal subalgebra by the boundary. Similar remarks
apply for the crosscap. From this we then conclude that the tadpoles are only defined when the left-
and rightmoving representations of $|\phi_{ij}\rangle$ are equal\footnote{More presicely, since
left-/right-movers can be viewed as in/outcoming states, the tadpoles for diagonal gluing
are only defined for representations that
are
each others charge conjugate. Alternatively, one can also consider tadpoles for charge conjugation gluing
conditions. Then the tadpoles are defined for diagonal closed strings. It is the latter formulation
that we are considering in this talk.}. So
\beq
\langle B |  \phi_{ij} \rangle \sim \delta_{ij} \;\;\; , \;\;\; \langle C |  \phi_{ij} \rangle
\sim \delta_{ij}\;\;\;.
\eeq
For a given closed oriented string spectrum described by a modular invariant $Z_{ij}$,
 we only need to know the tadpoles for a subset of primaries, namely those that couple diagonally. A primary $i$ for which $Z_{ii} \neq 0$ is called an {\em Ishibashi label}. Note that Ishibashi
 labels are degenerate when $Z_{ii} \geq 2 $. For simple current invariants, this can only happen when the
 currents have fixed points, so we will ignore this degeneracy problem (and its solution~\cite{FOE}) in this talk.
A second constraint only applies for the boundaries and is refered to as {\em
completeness}~\cite{completeness, DiZ}. It states: the
number of inequivalent boundaries of a given gluing condition equals the number of
Ishibashi labels. We then write
\beq
\langle B |  \phi_{ii} \rangle = \sum_a \cN_a \langle B_a |  \phi_{ii} \rangle
\eeq
where $|B_a\rangle$ is the boundary state of type $a$. The numbers $\cN_a$
are {\em Chan-Paton} factors and count how many times `brane' $a$ exists in the
theory. It is convenient to define the boundary and crosscap {\em coefficients}
\beq
B_{ia} = \langle B_a |  \phi_{ii} \rangle \;\;\;,\;\;\; \G_i = \langle C |  \phi_{ii} \rangle \;\;\; .
\eeq
A second constraint on the tadpoles is {\em integrality} and arises as follows. It is well known that the
spectrum of a string theory is encoded in the one-loop partition function. For a string theory including
worldsheet boundaries and crosscaps,
the total one-loop partition function has three additional contributions besides the torus $\cZ$:
 Klein bottle $\cK$,
 annulus $\cA$ and M\"obius strip $\cM$. In terms of characters we can write
 \beq
 \cK = \sum_m K_m \chi_m \;\;\;,\;\;\; \cA = \sum_{mab} \cN_a\cN_b A_{mab} \chi_m \;\;\;,\;\;\;
 \cM = \sum_{ma} \cN_a
 M_{ma} \hat{\chi}_m \;\;\; .
 \eeq
where the definition of $\hat{\chi}$ is such that even/odd level descendants of $m$
are
symmetrized/\newline
antisymmetrized when $M_{ma}>0$ and vice versa when $M_{ma}<0$.
The Klein bottle (anti)symmetrizes sector $Z_{mm}$ when $K_m$ is $(-)1$.
In order to have nonnegative state multiplicities in both the open and
 closed sector it is sufficient to have
\beq
 \hlf (Z_{mm} + K_m) \in \ZZ^+ \;\;\; , \;\;\; \hlf (A_{maa} + M_{ma}) \in \ZZ^+
 \;\;\; , \;\;\; A_{mab} = A_{mba} \in \ZZ^+ \;\;\; .
\eeq
where $\ZZ^+$ denote the positive integers including $0$.
It is intuitively clear that integrality constrains the tadpoles, since $\cA$ and $\cK$ contains two
boundaries and two crosscaps respectively and $\cM$ contains a boundary and a crosscap. We
require~\cite{cardy,SaSt}:
\beq
K_m = \sum_i S_{im} \G_i\G_i \;\;\; , \;\;\; A_{mab} = \sum_i S_{im} B_{ia} B_{ib}
\;\;\; , \;\;\;M_{ma} = \sum_i P_{im} \G_i B_{ia} \;\;\; .
\eeq
where $P=\sqrt{T}ST^2S\sqrt{T}$~\cite{P} and $S,T$ are the modular matrices of the chiral
 algebra under consideration.
 The third constraint is {\em tadpole cancellation}. All one-loop diagrams are finite
 when
 \beq \label{eq:tadpole}
\sum_a \cN_a B_{0a} = 2^{d/2} \G_0 \;\;\; , \;\;\; \sum_a \cN_a B_{ia} = -2^{d/2} \G_i
 \eeq
  for all Ishibashi primaries $i$ with $h_i=1$. Here $d$ is the uncompactified
  dimension of the fermionic string. Tadpole cancellation determines the gauge group in the
  open sector. We will require that $A_{0ab}$ is an involution on the boundary labels~\cite{SaSt}.
  This gauge group depends on the
  M\"obius coefficient $M_{0a}$ as follows. When $M_{0a}=+1$ the gauge group is $SO(\cN)$,
  when $M_{0a}=-1$ the gauge group is $Sp(\cN)$. Unitary gauge groups arise as follows.
  When $M_{0a}=0$,
  integrality in the open sector implies $A_{0aa}=0$ as well. Therefore, there must be a
  {\em conjugate} boundary label $\bar{a} \neq a$ for which $A_{0a\bar{a}}=1$. The conjugate
  pair represents a $U(\cN_a=\cN_{\bar{a}})$ gauge group.

We~\cite{FOE} have found a universal formula for the
boundary and crosscap coefficients in case the closed oriented string spectrum
 is described
by an {\em arbitrary} symmetric simple current invariant for {\em any} rational CFT.
 In case of a $\ZZ_N$ simple current group
without fixed points we have for the boundary~\cite{cardy,FSclass,FSorb} 
and crosscap coefficients~\cite{P,nondia,T}
\beq \label{eq:sol}
B_{i[j]} = \sqrt{N} \frac{S_{ij}}{\sqrt{S_{iK}}} \;\;\;,
\;\;\; \G_i = \frac{\sqrt{N}}{2} \frac{[\sigma(0) P_{iK} + \sigma(1) P_{i,JK}]}
{\sqrt{S_{iK}}}
\;\;\;.
\eeq
where the Ishibashi label $i$ is such that
$$Q_J(i) \equiv h_i +h_J - h_{Ji} \in \ZZ$$
and the boundary labels
$[j] = \{k=J^\a j|\a=1,...,N\}$  are given by simple current `orbits'. From~\cite{simple, intril}
\beq
S_{i,Jj}=e^{2\pi\i Q_J(i)}S_{ij}
\eeq
it follows
that the boundary coefficient does not depend on the orbit representative.
It is an easy exercise to show that the boundary coefficient has a left- and right inverse,
thus satisfying completeness.
The crosscap coefficient contains two
signs $\s(0)$ and $\s(1)$. When $N$ is odd, these signs must be such that $\G_i = \sqrt{N} \sigma(0) P_{iK}$.
When $N$ is even, these signs are independent. The primary field $K$ is a simple current called
{\em Klein bottle current}~\cite{klein}.
We can always choose $K=0$. When nontrivial, it is
must be local with all order two currents $J\in \ZZ_N$, i.e. $Q_K(J)=0$. To yield
inequivalent theories,
it is necessary that $K \notin \ZZ_N$ and $K^2=0$.

The solutions~\R{sol} reproduce all known ten-dimensional orientifolds~\cite{prop, sugimoto} when
applied to $SO(16)$ level $1$.
The $S$ and $P$ matrices are given by
\beq
	S_{ij} = \hlf e^{2\pi \i Q_j(i)} \;\;\; , \;\;\; P_{ij} = \delta_{ij}
\eeq
where $i,j = O_{16},V_{16},S_{16},C_{16}$.
The role of the Klein bottle
current can be nicely illustrated in the OB theory.
The
simple current group that defines $\cZ_{0B}$ is trivial, so all primaries can be used as a Klein bottle
current.
The Klein bottle
partition functions are given by
\bea
\cK^{[O]}_{0B} & = & O_{16} + V_{16} + S_{16} + C_{16} \;\;\; , \;\;\;
\cK^{[V]}_{0B}  =  O_{16} + V_{16} - S_{16} - C_{16} \;\;\; ,\\
\cK^{[S]}_{0B} & = & O_{16} - V_{16} + S_{16} - C_{16} \;\;\; , \;\;\;
\cK^{[C]}_{0B}  = O_{16} - V_{16} - S_{16} + C_{16} \;\;\; .
\eea
By completeness, there are four boundary labels denoted by $O,V,S,C$. The annulus and M\"obius for
the trivial Klein bottle current are
\bea
\cA^{[0]}_{0B} & = & \(\cN^2_O + \cN^2_V + \cN^2_S + \cN^2_C \) O_{16} + 2\(\cN_O\cN_V + \cN_S\cN_C\)
V_{16}\\
& & 2\(\cN_O\cN_S + \cN_V\cN_C\) S_{16} + 2\(\cN_O\cN_C + \cN_V\cN_S\) C_{16}
 \\
\cM^{[0]}_{0B} & = & \sigma(0) \(\cN_O + \cN_V + \cN_S + \cN_C\) \hat{O}_{16}
\eea
Tadpole cancellation~\R{tadpole} has to be imposed on $i=O_{16},C_{16},S_{16}$
and requires $\sigma(0)=1$ and an $[SO(\cN)\times SO(32-\cN)]^2$ gauge group. The effect of a nontrivial
Klein bottle current $K$ in the open sector can be summarized as follows:
\beq
A^{[K]}_{mab} = A^{[0]}_{Km,ab} \;\;\;\; , \;\;\; M^{[K]}_{Ka} = e^{2\pi\i Q_K(a)} M^{[0]}_{0a}
\eeq
where we only displayed the nonzero M\"obius coefficient. The only field that flows in the M\"obius is
the Klein bottle current. From~\R{PSSR}, we see that space-time vectors come from $O_{16}$, and therefore
a non-trivial Klein bottle leads to unitary gauge groups. The tadpole condition only has non-trivial
solutions when we relax the condition for $O_{16}$ (dilaton tadpole). For $K=V_{16}$ the gauge group is
$U(\cN) \times U(\cM)$. For the $K=S_{16}$ case we impose in addition
\beq
\cN_0=\cN_S \;\;\; , \;\;\; \cN_V=\cN_C
\eeq
since they are conjugate.
Tadpole cancellation then allows a $U(\cN)\times U(\cN - 32)$ gauge
group where $\cN \geq 32$. Strictly speaking, we have {\em two} such theories, labelled by
the free sign $\s(0)$.

The type OA theory is a $\ZZ_2$ simple current invariant based on the half-integer spin current $V_{16}$.
There are no Klein bottle currents. By completeness, we have two boundaries $[O]$
and $[S]$. The partition functions of this orientifold are
\bea
\cK_{0A} & = & O_{16} + V_{16} \;\;\; ,\\
\cA_{0A} & = & \(\cN_{[O]}^2 + \cN_{[S]}^2\) \(O_{16} + V_{16}\) + 2\cN_{[O]}\cN_{[S]}
\(S_{16} + C_{16}\)
\;\;\; ,\\
\cM_{0A} & = & \sigma(0) \(\cN_{[O]} + \cN_{[S]}\) \hat{O}_{16} + \sigma(1) \(\cN_{[O]} -
\cN_{[S]}\)  \hat{V}_{16} \;\;\; .
\eea
The dilaton tadpole is cancelled when $\s(0)=1$ and the gauge group equals $SO(\cN)\times
SO(32-\cN)$. The sign $\s(1)$ is unconstrained.

Note that the OA orientifold has tachyons in the open sector. Indeed,
from the truncation rules~\R{PSSR}, we see that $V_{16}$
contains the tachyon and there is no tadpole free
choice of $\cN$ to get rid of it. This happens also for the type OB orientifolds,
except when the Klein bottle current $K=S_{16}$ and the gauge group is $U(32)$.
From the Klein
bottle partition function it also follows that the closed string tadpole is projected
out~\cite{prop, BlFL}.

The type IIB string is a $\ZZ_2$ simple current invariant based on $S_{16}$.
The partition functions are
\bea
\cK_{IIB} & = &  O_{16} + S_{16} \;\;\; ,\\
\cA_{IIB} & = & \(\cN_{[O]}^2 + \cN_{[V]}^2\) \(O_{16} + S_{16}\) + 2\cN_{[O]}\cN_{[V]} \(V_{16} +
C_{16}\) \;\;\; ,\\
\cM_{IIB} & = & \sigma(0) \(\cN_{[O]} + \cN_{[V]}\) \hat{O}_{16} + \sigma(1)
\(\cN_{[O]} - \cN_{[V]}\) \hat{S}_{16} \;\;\; .
\eea
Note that these expressions are the same as those of the type OA descendant when $S$ and $V$ are
interchanged. The tadpole cancellation conditions for the dilaton and (unphysical) axion are
\beq
\cN_{[O]} + \cN_{[V]} = \s(0) 32 \;\;\; , \;\;\; \cN_{[O]} - \cN_{[V]} = -\s(1) 32 \;\;\; .
\eeq
When we insist on dilaton tadpole cancellation, we must choose $\s(0)=1$.
 When $\s(1)=1$ as well, we have a $SO(32)$ gauge group from $32$ branes of type
$[V]$. For $\s(1)=-1$ we have a $SO(32)$ gauge group from the $[O]$ branes. We can identify the
signs $-\s(0)$ and $\s(1)$ with the tension and RR charges of an O-plane\footnote{The tension
is $-\s(0)$ since the ten-dimensional
dilaton appears at the first excited level of $O_{16}\bar{O}_{16}$.}.
Similarly, we can identify the
$[O]$-boundary states as D-branes and the $[V]$-boundary states as anti-D-branes.
 Note also that when
we relax the
dilaton tadpole condition, we can choose $\s(0)=-1$ and thus allow symplectic gauge groups
$Sp(\cN)\times Sp(32-\cN)$. For $Sp(32)$ the open string tachyon is removed from
 the spectrum~\cite{sugimoto}.

\section{Conclusions}

We have rediscovered all ten dimensional orientifolds using simple currents. These theories can be
supersymmetric or nonsupersymmetric, may have dilaton tadpoles and/or tachyon instabilities. It is
intriguing to realize that the fermionic string theories in ten
dimensions are closely related to the $\ZZ_2 \times \ZZ_2$ centre of $SO(16)$.
 (See also the contribution of Laurent Houart to these proceedings~\cite{houart}.)

We have found universal formulas for boundary and crosscap coefficients for {\em all} symmetric
simple current invariants for {\em any} rational chiral algebra. These formulas basically
contain
the $S$ and $P$ matrices of a chiral algebra and a few signs and phases, which are known for
many RCFT's, like WZW models and cosets. Our construction can therefore be used to compute the spectra, brane
tensions etcetera for a {\em huge} number of string theories~\cite{kac}.  

We believe that the classification of D-branes and O-planes in rational CFT's is a first step
in understanding the complete spectrum of branes and planes at arbitrary points in the moduli space
of string theories. There has been a lot of interest in the comparison of D-branes at rational and
large volume points in moduli spaces of Calabi-Yau compactifications~\cite{BrDLR}.
 The hope is that
 one can derive from this the brane spectrum at arbitrary modulus. A powerful tool is mirror symmetry.
Under mirror symmetry, so-called A-type branes and B-type branes are interchanged and each
can be used to probe a different part of the moduli space. In this respect it is interesting to note
that the distinction between mirror compactifications and therefore A- and B-type branes
is another application of simple currents~(see for instance \cite{FSW}).

\bigskip

{\bf Acknowledgements:}
I would like to thank Bert Schellekens for many stimulating discussions.
I am grateful to the organizers of the EC-TMR workshop
 for giving me the opportunity to present my work.

\providecommand{\href}[2]{#2}\begingroup\raggedright\endgroup

\end{document}